\documentstyle[12pt]{article}

\begin{document}

\setcounter{page}{1}

\pagestyle{plain} \vspace{1cm}
\begin{center}
\Large{\bf Cosmological aspects of a vector field model}\\
\small \vspace{1cm}
{\bf S. Davood Sadatian}\footnote{sd-sadatian@um.ac.ir}\\
\vspace{0.5cm} {\it Department of Physics,
Faculty of Basic Sciences,\\
University of Neyshabur,\\
P. O. Box 91136-899, Neyshabur, Iran
 }\\
\end{center}
\vspace{1.5cm}
\begin{abstract}
We have studied a DGP-inspired braneworld scenario where the idea of
Lorentz invariance violation has been combined into a specifying
preferred frame that embed a dynamical normal vector field to brane.
We propose the Lorentz violating DGP brane models with enough
parameters can explain crossing of phantom divide line. Also we have
considered the model for proper cosmological evolution that is
according to the observed behavior of the equation of state. In
other view point, we have described a Rip singularity
solution of model that occur in this model. \\
\\
{\bf PACS}: 04.50.+h, 98.80.-k\\
{\bf Key Words}:  DGP Brane Cosmology, Braneworld Cosmology,
Scaler-Vector-Tensor Theories, Lorentz Invariance Violation.
\end{abstract}
\vspace{1.5cm}
\newpage

\section{Introduction}
According observation data from WMAP, universe accelerated expanding
at the current epoch [1,2,3,4,5]. These astrophysical observations
also show that the universe is spatially, and is composed of about
$27\%$ of dark matter, and $73\%$ of homogeneously distributed new
type of negative pressure matter, known as dark energy, that lead to
the current accelerated expansion in universe. From another view
point, theories of extra dimensions, in which the observed universe
is realized as a brane embedded in a higher dimensional spacetime,
have attracted some attentions. Particularly, the model proposed by
Dvali, Gabadadze and Porrati (DGP) [6,7,8] is different from others.
In this regard, DGP model also predicts deviations from the standard
4-dimensional gravity in over large distances. However, impact of
Lorentz invariance violating models (LIV) on cosmology has been
studied in [9,10,11]. This models has been used in the context of
scalar-vector-tensor theories [9,11]. One important observation has
been made in reference [11,12] which considered accelerated
expansion and crossing of phantom divide line with one minimally
coupled scalar field in the attendance of a Lorentz invariance
violating vector field. As one has shown in [13], quintessence model
with a single minimally coupled scalar field has not the capability
to describe crossing of phantom divide line, $\omega=-1$. But, a
single non-minimally coupled scalar field is suitable to cross the
phantom divide line [14].

In following, we study a new method of performing local Lorentz
violation in a gravitational setup to make the existence of a tensor
field with a non-vanishing expectation value, and then add this
tensor to gravity or matter and scalar fields. The example of this
approach is to used a single time-like vector field with unit norm.
This vector field choose a preferred frame at each point in
space-time and a matter field coupled to it will endure a violation
of local Lorentz invariance. A method of this model was introduced
Kostelecky and Samuel in [15]. Here we choose the vector to have
dynamics, and fixed norm in the action.

Our procedure in this paper is deriving the basic equations of
motion for the most general theory of a time-like vector field
$u^\mu$ with an action $S=S_{Bulk}+S_{Brane}$ where
$S_{Brane}=S_{EH}+S_{\phi}+S_{m}+S_{u}$ [11]. Then with numerical
calculations, we study some cosmological aspects in this model in
present the vector field in our formalism. In this regard, we study
other solutions admitted as Rip singularity , that occur in the
condition $\omega < -1 $ increases rapidly. However, it possible
different types of singularity, depending energy density and scale
factor how increases with time[16-19,31-36]. In following we just
focus on Big or Little singularity and other cases of singularity
maybe appear in this model with tuning parameters space.

\section{A Lorentz Violating DGP Brane Model}
As we shown in[11], for study Lorentz violation, we assume a vector
$u^\mu$ in the extra dimension. So, this local frame at space-time
is unavoidably selected as the preferred frame. However the action
of the Lorentz violating DGP scenario in the presence of a minimally
coupled scaler field and vector field on the Brane can be written as
the sum of two separate parts
\begin{equation}
S=S_{Bulk}+S_{Brane},
\end{equation}
where $S_{Bulk}$\, and\, $S_{Brane}=S_{EH}+S_{\phi}+S_{m}+S_{u}$ are
\begin{equation}
S_{Bulk}=\int d^{5}x\frac{m^{3}_{4}}{2}\sqrt{-g}{\cal R},
\end{equation}
$$S_{Brane}=\Bigg[\int d^{4}x\sqrt{-q}\bigg(\frac{m_{3}^{2}}{2}
R[q]-\frac{1}{2} q^{\mu\nu} \nabla_{\mu}\phi\nabla_{\nu}\phi
-V(\phi) + m^{3}_{4}\overline{K}+ {\cal{L}}_{m}+$$
\begin{equation}
[-\beta_1 \nabla^\mu u^\nu \nabla_\mu u_\nu-\beta_2 \nabla^\mu u^\nu
\nabla_\nu u_\mu -\beta_3 \left( \nabla_\mu u^\mu \right)^2-\beta_4
u^\mu u^\nu \nabla_\mu u^\alpha \nabla_\nu u_\alpha+\lambda \left(
u^\mu u_\mu +1 \right)]\bigg)\Bigg]_{y=0}.
\end{equation}
\, $y$ is coordinate of the fifth dimension, $m_4^3$ point to the
constant in the Bulk and $m_3^2$ used for Brane, for more details
see [11]. We consider brane is located at $y=0$.\,$g_{AB}$ is five
dimensional bulk metric with Ricci scalar ${\cal{R}}$, also
$q_{\mu\nu}$ is induced metric on the brane with induced Ricci
scalar $R$.\, $g_{AB}$ and $q_{\mu\nu}$ are connected via
$q_{\mu\nu}={\delta_{\mu}}^{A}{\delta_{\nu}}^{B}g_{AB}$.\, This
action is suitable for add any non-gravitational degrees of freedom
in the model of Lorentz violating scalar-tensor-vector theory. In
other hand, we take $u^\mu u_\mu = -1$ and the expectation value of
vector field $u^\mu$ is $<0| u^\mu u_\mu |0> = -1$\,[20].
$\beta_i(\phi)$ ($i=1,2,3,4$) are temporary parameters with
dimension of mass squared,\,$\lambda$ is also a Lagrange multiplier
field. Note that $\sqrt{\beta_{i}}$ are mass scale of violating
Lorentz invariance [9,20,21].\\
A motivation for choosing the vector action in term (3) discussed in
Ref. [20]. Also cosmological consequences of this action are studied in Ref.[9,11,21].\\
In the action (3) ordinary matter part is shown by Lagrangian
${\cal{L}}_{m}\equiv {\cal{L}}_{m}(q_{\mu\nu},\psi)$ where $\psi$ is
matter field and its energy-momentum tensor is[22]
\begin{equation}
T_{\mu\nu}=-2\frac{\delta{\cal{L}}_{m}}{\delta
q^{\mu\nu}}+q_{\mu\nu}{\cal{L}}_{m}.
\end{equation}
The scalar field Lagrangian,\, ${\cal{L}}_{\phi}=-\frac{1}{2}
q^{\mu\nu} \nabla_{\mu}\phi\nabla_{\nu}\phi -V(\phi)$,\,\, gives the
following energy-momentum tensor
\begin{equation}
 \tau_{\mu\nu}=\nabla_\mu\phi\nabla_\nu\phi-\frac{1}{2}q_{\mu\nu}(\nabla\phi)^2
-q_{\mu\nu}V(\phi).
\end{equation}
The energy-momentum tensor of vector field also is by usual
formulate
\begin{equation}
T_{\mu\nu}^{(u)}=-2\frac{\delta{\cal{L}}^{(u)}}{\delta
q^{\mu\nu}}+q_{\mu\nu}{\cal{L}}^{(u)}.
\end{equation}
The Bulk-brane Einstein's equations obtained from action (1) are
given by
\begin{equation}
m^{3}_{4}\left({\cal R}_{AB}-\frac{1}{2}g_{AB}{\cal R}\right)+
m^{2}_{3}{\delta_{A}}^{\mu}{\delta_{B}}^{\nu}\left(R_{\mu\nu}-
\frac{1}{2}q_{\mu\nu}R\right)\delta(y)=
{\delta_{A}}^{\mu}{\delta_{B}}^{\nu}\Upsilon_{\mu\nu}\delta(y)
\end{equation}
where $\Box^{(4)}$ is 4-dimensional(brane) d'Alembertian and
$\Upsilon_{\mu\nu}=T_{\mu\nu}+\tau_{\mu\nu}+T_{\mu\nu}^{(u)}$\,.\\
From equation (7) we find
\begin{equation}
G_{AB}={\cal R}_{AB}-\frac{1}{2}g_{AB}{\cal R}=0
\end{equation}
and
\begin{equation}
G_{\mu\nu}=\left(R_{\mu\nu}-
\frac{1}{2}q_{\mu\nu}R\right)=\frac{\Upsilon_{\mu\nu}}{m^{2}_{3}}
\end{equation}
for bulk and brane respectively. We use the following line element
to derive cosmological equations of our model
\begin{equation}
ds^{2}=q_{\mu\nu}dx^{\mu}dx^{\nu}+b^{2}(y,t)dy^{2}=-n^{2}(y,t)dt^{2}+
a^{2}(y,t)\gamma_{ij}dx^{i}dx^{j}+b^{2}(y,t)dy^{2}
\end{equation}
we use $n(0,t)=1$ and ${\cal{N}}=1$ after
the variation [11]. \\
Now we can obtain components of the total energy-momentum tensor as
follow
\begin{equation}
\rho=\rho_m+\rho_u+\rho_\phi
\end{equation}
and
\begin{equation}
p=p_m+p_u+p_\phi
\end{equation}
Here we take the energy-momentum tensor for the matter as a perfect
fluid with energy density $\rho_m$ and pressure $p_m$ is
$T_{\mu\nu}=(\rho_m+p_m)N_\mu N_\nu+p_m q_{\mu\nu}$, where $N_\mu$
is a unit time-like vector field illustrate the fluid four-velocity.
We also assume a  usual equation of state for the fluid as
$p_m=(\gamma-1)\rho_m$ that $1\leq\gamma\leq 2$. Energy density and
pressure of minimally coupled scalar field are taken as
\begin{equation}
\rho_{\phi}=\left[\frac{1}{2}\dot{\phi}^{2}+n^{2}V(\phi)\right]_{y=0},
\end{equation}
and
\begin{equation}
p_{\phi}=\left[\frac{1}{2n^{2}}\dot{\phi}^{2}-V(\phi)\right]_{y=0},
\end{equation}
where a dot point to the derivative respect to cosmic time\, $t$.
The stress-energy for the vector field also assume the form of a
perfect fluid, with an energy density determined by [10]
\begin{equation}
\rho_u=-3\beta H^2
\end{equation}
and a pressure
\begin{equation}
p_u=\beta
H^2\Big[3+2\frac{\dot{H}}{H^2}+2\frac{\dot{\beta}}{H\beta}\Big]
\end{equation}
where we have assume the parameter
$\beta\equiv(\beta_1+3\beta_2+\beta_3)$ and
$H=\frac{\dot{a}(0,t)}{a(0,t)}$ is Hubble parameter. In the absence
of vector field, all $\beta_i = 0$, In following, we stress for
obtain equation of state in general case one should varying the
action (1), that It has been calculated in Ref[9,10 11,12,15,20, 21
and 22]. Here we just used their results and summarized
calculations. Hence, we write the general equation of state which
takes the following form
\begin{equation}
\omega=\Big[\frac{p_m+\beta
H^2\Big[3+2\frac{\dot{H}}{H^2}+2\frac{\dot{\beta}}{H\beta}\Big]
+\frac{1}{2n^{2}}\dot{\phi}^{2}-V(\phi)}{\rho_m-3\beta
H^2+\frac{1}{2}\dot{\phi}^{2}+n^{2}V(\phi)}\Big]_{y=0}
\end{equation}
and for obtain generalized Friedmann equations, we use the effective
Einstein equation (for more details see[11,23]), therefore we have
\begin{equation}
3\bigg(H^{2}+\frac{k}{a^{2}}\bigg)={{\cal{E}}^{0}}_{0}+
\frac{1}{12m_{4}^{6}}\bigg[\rho_m-3\beta
H^2+\frac{1}{2}\dot{\phi}^{2}+n^{2}V(\phi)-3m_{3}^{2}\Big(H^{2}+\frac{k}{a^{2}}\Big)\bigg]^{2}.
\end{equation}
As we calculated in above equations, there are many parameters, so
we expect the Lorentz violating brane model, suitable for study the
current stage acceleration of the universe and other cosmological
aspects with fine tuning parameters.
\section{Setup for useful equations}
In following, we consider scalar equation of state $\omega_\phi(t)$
which is a candidate for the dark energy in future discussions. In
this regard, the equation of state parameter on the brane given by
\begin{equation}
\omega_\phi=\frac{p_\phi}{\rho_\phi}=\Big[\frac{\frac{1}{2n^{2}}\dot{\phi}^{2}-V(\phi)}
{\frac{1}{2}\dot{\phi}^{2}+n^{2}V(\phi)}\Big]_{y=0}
\end{equation}
Now we obtain scalar field of equation of state in two ways. First,
we have Friedmann equation from equation (18) as
\begin{equation} H^{2}=\frac{{{\cal{E}}}_{0}}{3a^4}+
\frac{1}{36m_{4}^{6}}\bigg[-3\beta
H^2+\frac{1}{2}\dot{\phi}^{2}+V(\phi)-3m_{3}^{2}H^{2}\bigg]^{2}
\end{equation}
where $k=0$ and we assume non matter contain on the model that means
$\rho_m=0$ and $p_m=0$, where
$\dot{{{\cal{E}}}^{0}}_{0}+4H{{\cal{E}}^{0}}_{0}=0$ \,\,and
${{\cal{E}}^{0}}_{0}=\frac{{{\cal{E}}_{0}}}{a^{4}}$ with
${\cal{E}}_{0}$ as an integration constant[23]. Then we obtain
dynamic of scalar field from above equation
\begin{equation}
\dot{\phi}^2=6H^2(\beta+m_3^2)-2V(\phi)+12m_4^3\epsilon\sqrt{H^2+\frac{{{\cal{E}}}_{0}}{3a^4}}
\end{equation}
where $\epsilon=\pm1$ has main role in next calculations. For
determine dynamic of scalar field, we use to a well-known potential:
models with potential of the form \, $V(\phi)=\lambda'\phi^2$\,[24].

Second, Such as we see equation (21) depend on potential functions,
therefore we will obtain the scalar field of equation of state in
our model that independent of potentials. So the energy equation for
vector field taken as
\begin{equation}
\dot{\rho}_u+3H(\rho_u+p_u)=+3H^2\dot{\beta}
\end{equation}
and for the scalar field
\begin{equation}
\dot{\rho}_\phi+3H(\rho_\phi+p_\phi)=-3H^2\dot{\beta}.
\end{equation}
The total energy equation in presence of both the vector and the
scalar fields is
\begin{equation}
\dot{\rho}+3H(\rho+p)=0,~~~ (\rho=\rho_u+\rho_\phi)
\end{equation}
Now we obtain dynamic of scalar field by differentiating equation
(13) with respect to $t$ and using equation (23) then we have
\begin{equation}
\ddot{\phi}+3H\dot{\phi}+3H^2\beta_{,\phi}+V_{,\phi}=0
\end{equation}
and by differentiating equation (20) with respect to $t$ and using
equation (25) we have
\begin{equation}
\dot{\phi}=\mp2m_4^3(\frac{2H_{,\phi}}{\dot{\phi}H}+\frac{12{\cal{E}}_{0}}{\dot{\phi}^2\dot{a}a^3})^{\frac{1}{2}}
-2H\beta_{,\phi}-2\beta H_{,\phi}-m_3^2\frac{H_{,\phi}}{H}
\end{equation}
where we assume $ H $ and $\beta $ depended on $\phi$ . For
simplicity in following we take ${\cal{E}}_{0}=0$.\\
In other hand, with substituting Equation (26) into the Friedmann
equation (20) obtain the potential of the scalar field as
\begin{equation}
V(\phi)=\frac{1}{2}\frac{12H+6{
m_3}^{2}{H}^{2}{m_4}^{3}-{\dot{\phi}}^{2}{{m_4}^{3}+6{H}^{2}{m_4}^{3}\beta}}{{
m_4}^{3}}
\end{equation}
and
\begin{equation}
V(\phi)=\frac{1}{2}\frac{-12H+6{
m_3}^{2}{H}^{2}{m_4}^{3}-{\dot{\phi}}^{2}{{m_4}^{3}+6{H}^{2}{m_4}^{3}\beta}}{{
m_4}^{3}}
\end{equation}
Also factor $\dot{\phi}$ from equation (26) give as
 $$\dot{\phi}={ 1+2\beta_{,\phi}\beta H_{,\phi}+
 .6\beta_{,\phi}H_{,\phi}+ .6\beta H}\times$$
\begin{equation}
 \frac{\sqrt[3]{\Big({
 10H_{,\phi}+{4}+{4}\beta H_{,\phi}+{5}\beta_{,\phi}+{4}
\beta_{,\phi}\beta+{80}\beta H+A}\Big)}}{H}
\end{equation}
$$ \times\cal{O}(\phi)$$
where $A={2}\sqrt{100+10H_{,\phi}+30\beta+
 40\beta_{,\phi}+30\beta_{,\phi}\beta+10\beta H}
\,\epsilon'H$ and $\epsilon'=\pm$ 1. We emphasize that equation (26)
have several solutions, here we choose a real solution of them.
Besides, equation (29) has not the analytical solution that we
exactly use it for cosmological consideration. Hence, we just study
some phenomenological aspects from Figure 1 and choose a best fit
scalar filed with equation (29) for analytical consideration.
\begin{figure}
\begin{center}\includegraphics{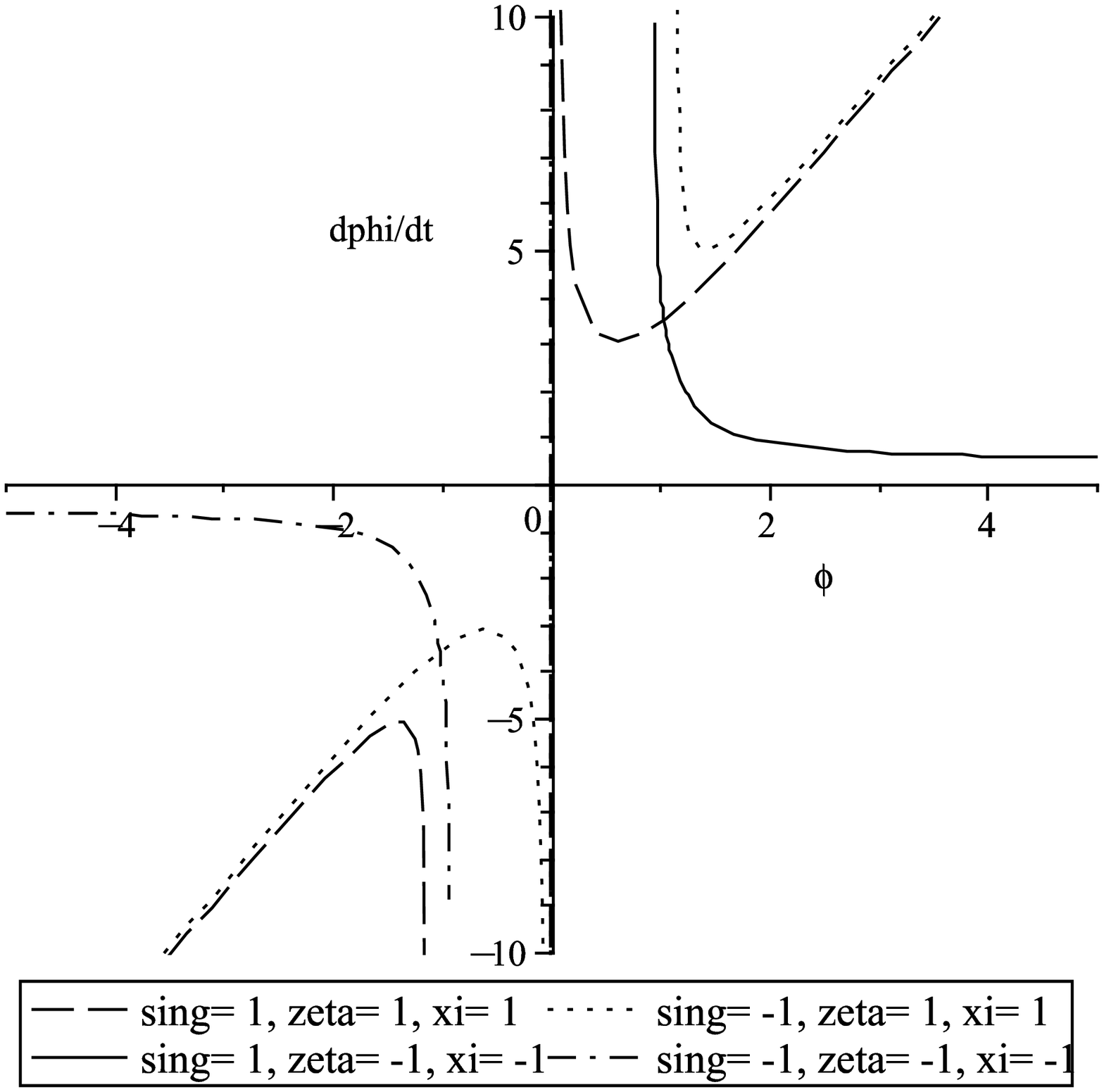} \vspace{5.5cm}
\end{center}
\caption{\small {Variation of~ $\frac{d\phi}{dt}$ relative to $\phi$
in equation (29) for different values of $\xi$ and $\zeta$ according
ansatz $H=H_0\phi^{\zeta}~ and~ \beta(\phi) = m\phi^\xi$.}}
\end{figure}\\
In follows, we calculate scalar field of equation of state with
using equation (20). Therefore we have two equations for $H^2$ as
\begin{equation}
H^2=\frac{1}{3}\,{\frac
{\rho_{\phi}\,{m_3}^{2}+\beta\,\rho_{\phi}+6\,{m_4}^{6}+2\,\sqrt
{3\,{m_4}^{6}
\rho_{\phi}\,{m_3}^{2}+3\,{m_4}^{6}\beta\,\rho_{\phi}+9\,{m_4}^{12}}}{2\,\beta\,{m_3}^{2}+{
\beta}^{2}+{m_3}^{4}}}
\end{equation}
and
\begin{equation}
H^2=\frac{1}{3}\,{\frac
{\rho_{\phi}\,{m_3}^{2}+\beta\,\rho_{\phi}+6\,{m_4}^{6}-2\,\sqrt
{3\,{m_4}^{6}
\rho_{\phi}\,{m_3}^{2}+3\,{m_4}^{6}\beta\,\rho_{\phi}+9\,{m_4}^{12}}}{2\,\beta\,{m_3}^{2}+{
\beta}^{2}+{m_3}^{4}}}.
\end{equation}
For each branch of $H^2$ means equation (30) and (31) with using
equation (23) we obtain equation
\begin{equation}
E1+E2(1+\omega_{\phi})=-\dot{\beta}E3
\end{equation}
where
\begin{equation}
E1=\frac{\dot{\rho_{\phi}}}{\rho_{\phi}}={\frac
{{\dot{H}}}{H}}+{\frac {{\dot{\beta}}\,H{{m_4}}^{3}+\beta
\,{\dot{H}}\,{{m_4}}^{3}+{{m_3}}^{2}{\dot{H}}\,{{m_4}
}^{3}}{2\,{\epsilon_4}+\beta\,H{{m_4}}^{3}+{{m_3}}^{2}H{{m_4}}^{3}}}
\end{equation}
and
\begin{equation}
E2=\frac{3\,{\epsilon_1}\, \left( {{m_4}}^{3}+{\epsilon_2}\,\sqrt
{{\frac {{{ m_4}}^{9}+{H}^{2} \left( \beta+{{m_3}}^{2} \right)
^{2}{{ m_4}}^{3}+2\,H{\epsilon_4}\, \left( \beta+{{m_3}}^{2} \right)
}{{{m_4}}^{3}}}} \right)} { \left( \beta+{{m_3}}^{2} \right)}
\end{equation}
and
$$E3=\frac {2\,{\epsilon_3}\,\sqrt {{{m_4}}^{3} \left(
{{m_4}}^{9} +{H}^{2} \left( \beta+{{m_3}}^{2} \right)
^{2}{{m_4}}^{3}+2 \,H{\epsilon_4}\, \left( \beta+{{m_3}}^{2} \right)
\right) }{{m_4}}^{3}}{H \left( {{m_4}}^{3} \left( \beta+{{m_3}}^ {2}
\right) H+2\,{\epsilon_4} \right) \left( \beta+{{m_3}}^{2}
 \right) ^{2}}$$

\begin{equation}
+\frac{{H}^{2} \left( \beta+{{m_3}}^{2} \right)
^{2}{{m_4}}^{3}+2\,H{\epsilon_4}\, \left( \beta+{{m_3}}^{2}\right)
+2\, {{m_4}}^{9}}{H \left( {{m_4}}^{3} \left( \beta+{{m_3}}^ {2}
\right) H+2\,{\epsilon_4} \right) \left( \beta+{{m_3}}^{2}
 \right) ^{2}}
\end{equation}
where $\epsilon_{1}=\epsilon_{2}=\epsilon_{3}=\epsilon_{4}=\pm 1$.
We can calculate $\omega_{\phi}$ after simplify from equation (32)
as

$$\omega_{\phi}=1/6\, \Bigg[ -6\,{\epsilon_2}\, \left(
\beta+{{m_3}}^{2} \right) H{ \epsilon_1}\, \left( 1/2\,{{m_4}}^{3}
\left( \beta+{{m_3}}^{2}
 \right) H+{\epsilon_4} \right)\times$$

 $$\sqrt {{\frac {{{m_4}}^{9}+{H}^{2}
 \left( \beta+{{m_3}}^{2} \right) ^{2}{{m_4}}^{3}+2\,H{\epsilon_4}\, \left( \beta+{{m_3}}^{2} \right)
 }{{{m_4}}^{3}}}}-$$

 $$2 \, {\beta_{,\phi}}\,{\dot{\phi}}\,{\epsilon_3}\,\sqrt
{{{m_4}}^{3}
 \left( {{m_4}}^{9}+{H}^{2} \left( \beta+{{m_3}}^{2}
 \right) ^{2}{{m_4}}^{3}+2\,H{\epsilon_4}\, \left( \beta+{{m_3
}}^{2} \right)  \right) }{{m_4}}^{3}-$$

$$3\,{{m_4}}^{3} \left( \beta+{{m_3}}^{2} \right) ^{2} \left(
2/3\,{\beta_{,\phi}}\,{\dot{\phi}}+{\epsilon_1}\,{{m_4}}^{3} \right)
{H}^{2}-$$

$$6\, \left(
 \left( 1/3\,{H_{,\phi}}\,{\dot{\phi}}\,{{m_3}}^{4}+2/3\,{H_{,\phi}
}\,{\dot{\phi}}\,\beta\,{{m_3}}^{2}+{\epsilon_1}\,{\epsilon_4}+1/3\,{
\beta}^{2}{H_{\phi}}\,{\dot{\phi}} \right)
{{m_4}}^{3}+1/3\,{\beta_{,\phi}}\,{\dot{\phi}}\,{\epsilon_4} \right)
\left( \beta+{{m_3}}^ {2} \right) H-$$

$$2\, \left( {{m_4}}^{9}{\beta_{,\phi}}+{H_{,\phi}}\,{ \epsilon_4}\,
\left( \beta+{{m_3}}^{2} \right) ^{2} \right) {\dot{\phi}} \Bigg]
\Bigg[\left( {{m_4}}^{3}+{\epsilon_2}\,\sqrt {{\frac {{{
m_4}}^{9}+{H}^{2} \left( \beta+{{m_3}}^{2} \right) ^{2}{{
m_4}}^{3}+2\,H{\epsilon_4}\, \left( \beta+{{m_3}}^{2} \right)
}{{{m_4}}^{3}}}} \right)^{-1}$$

\begin{equation}
\times\left( \beta+{{m_3}}^{2}
 \right) ^{-1}{H}^{-1}{{\epsilon_1}}^{-1} \left( 1/2\,{{m_4}}^{3}
 \left( \beta+{{m_3}}^{2} \right) H+{\epsilon_4} \right) ^{-1}\Bigg]
\end{equation}
In following we study our models by using above equations for obtain
some constraint of parameters that be according to observation data
and also discuss about other cosmological aspects.
\section{Fine tuning parameters }
In this section we consider our equation that obtained in above
section for determine some cosmological aspects. Therefore we need
to solve equation (36) to study crossing of phantom divide barrier
$\omega_{\phi}$ in model. In first stage we should obtain dynamics
of scalar field $\phi$ with equation (21). This will be achieved
only if the Hubble parameter $H(\phi(t))$ and the vector field
coupling,\, ${\beta}(\phi(t))$ are known. Hence, our strategy is to
choose some cases of the Hubble parameter $H(\phi(t))$ and the
vector field coupling $\beta(\phi(t))$ for considering possible
crossing of phantom divide barrier (PDL) and Rip singularity in this
model. Also we should obtain suitable domains of parameter space
which have the capability to explain Rip singularity and crossing of
phantom divide line by equation of state parameter.

In following, we take a general case of the vector field coupling
and the Hubble parameter that are functions of scalar field $\phi$

\begin{equation}
H=H_0\phi^{\zeta} \ , \quad  \beta(\phi) = m\phi^\xi
\end{equation}
where $H_0$ and $m$ are positive constant parameters. Here we used
anzats (37) because considered some cases of the solution and
verified the stability in previous works [12]. Also other authors
[10 and references therein] have been used anzats (37) for some
cosmological solutions, for example, late time acceleration,
deceleration parameters and etc. So let us using equation (21) for
obtain dynamic of scalar field equation with assuming case (37) as
\begin{equation}
\dot{\phi}=6{H_0}^2\phi^{2\zeta}(m\phi^{\xi}+{m_3}^2)-\lambda'\phi^2+12{m_4}^3\epsilon{H_0}\phi^{\zeta}
\end{equation}
for potential $V(\phi)=\lambda'\phi^2$.\\
If we use a simple case from a acceptable range of $\xi$ and $\zeta$
for analytical solution of scalar field equation (38), we obtain
analytical solution for dynamic of scalar filed equation as follows
$$\phi(t)
=\frac{1}{2}\Bigg[\Bigg(36{{m_4}}^{6}{\epsilon}^{2}{{H_0}}^{2}{e^{2{A_0}\sqrt
{6{{H_0}}^{2}{{m_3}}^{2}-\lambda'}}}+9{{H_0}}^{4}{m}^{2}{e^{2{A_0}\sqrt
{6{{H_0}}^{2}{{m_3}}^{2}-\lambda'}}}-$$

$$12{e^{\left(t+{A_0}\right)\sqrt{6{{H_0}}^{2}{{ m_3}}^{2}-\lambda'}
}}\sqrt{6{{H_0}}^{2}{{m_3}}^{2}-\lambda'}{{m_4}}^{3}\epsilon{H_0}+6{e^{2t\sqrt
{6\,{{H_0}}^{2}{{m_3}}^{2}-\lambda'}}}{{H_0}}^{2}{{m_3}}^{2}-$$

$${e^{2t\sqrt {6{{H_0}}^{2}{{m_3}}^{2}-\lambda'}}}\lambda'-6{e^{\left(t+{A_0}\right)\sqrt{6{{H_0}}^{2}{{m_3}}^{2}-\lambda'}}}
\sqrt{6{{H_0}}^{2}{{m_3}}^{2}-\lambda'}{{H_0}}^{2}m+36{{H_0}}^{3}m{{m_4}}^{3}\epsilon{e^{2{A_0}\sqrt
{6{{H_0}}^{2}{{m_3}}^{2}-\lambda'}}}
 \Bigg)$$
\begin{equation}
{e^{-\left(t+{A_0}\right)\sqrt{6{{H_0}}^{2}{{m_3}}^{2}-\lambda'}
}}\Bigg]\Big[\left(6{{H_0}}^{2}{{m_3 }}^{2}-\lambda'\right)
^{3/2}\Big]^{-1}
\end{equation}
where we take $\xi=-1$ , $\zeta=1$ and $A_0$ is an integration constant.\\
In the following we consider equation of state (36) with using
equation (39) for exact dynamic of equation of state, Then study
crossing phantom divided line and other options in our model. The
result of our numerical calculation has been shown in figures 2-5
and 6-9. The summarize of results given in table 1. According
figures 2-5 and table 1, Our model can explain crossing phantom
divided, that is relevant to observational data.
\begin{figure}[htp]
\includegraphics{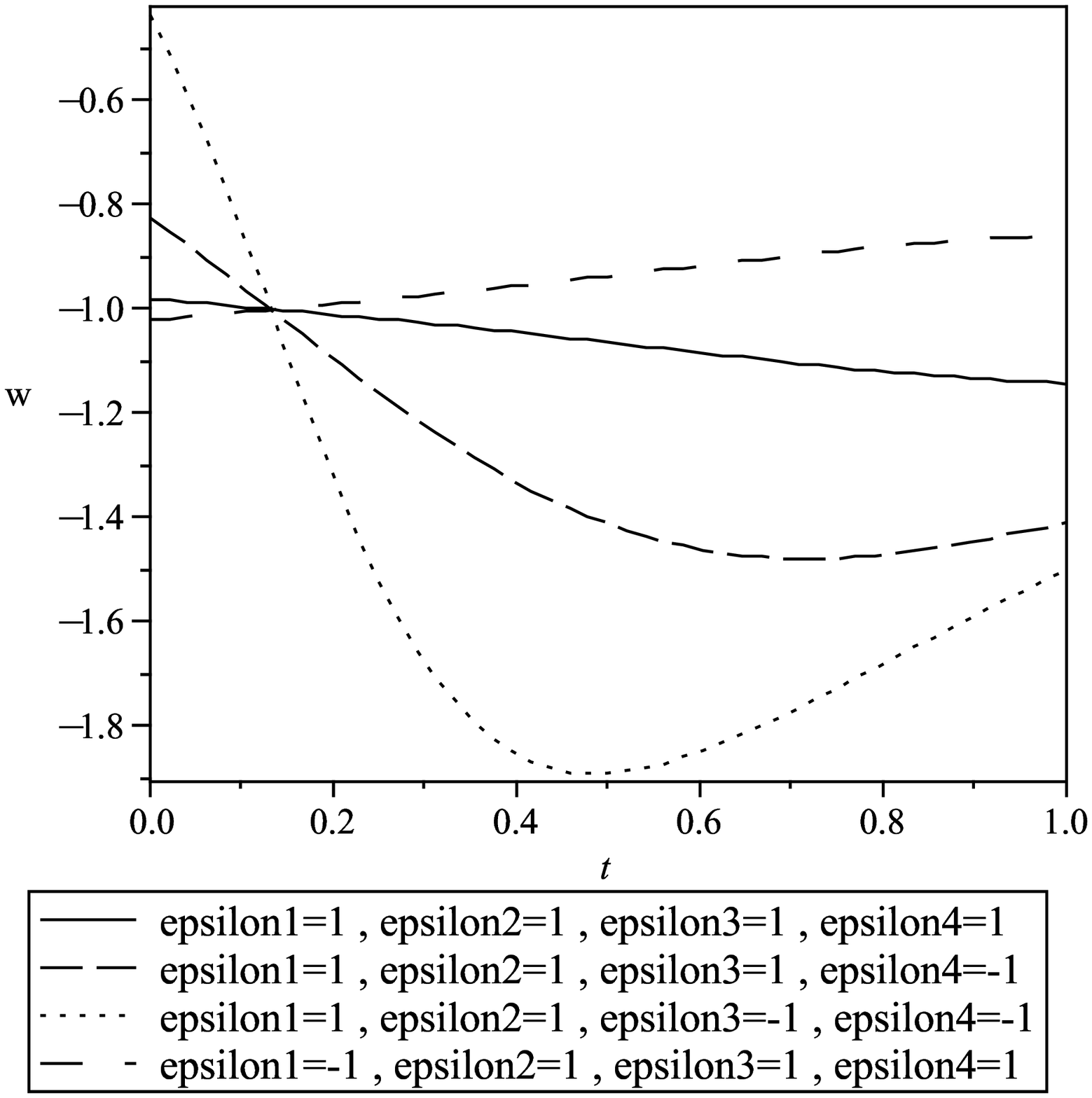} \vspace{7 cm}
 \caption{\small {Crossing of phantom divide line. }
 \hspace{11cm}}
 \vspace{10cm}

\includegraphics{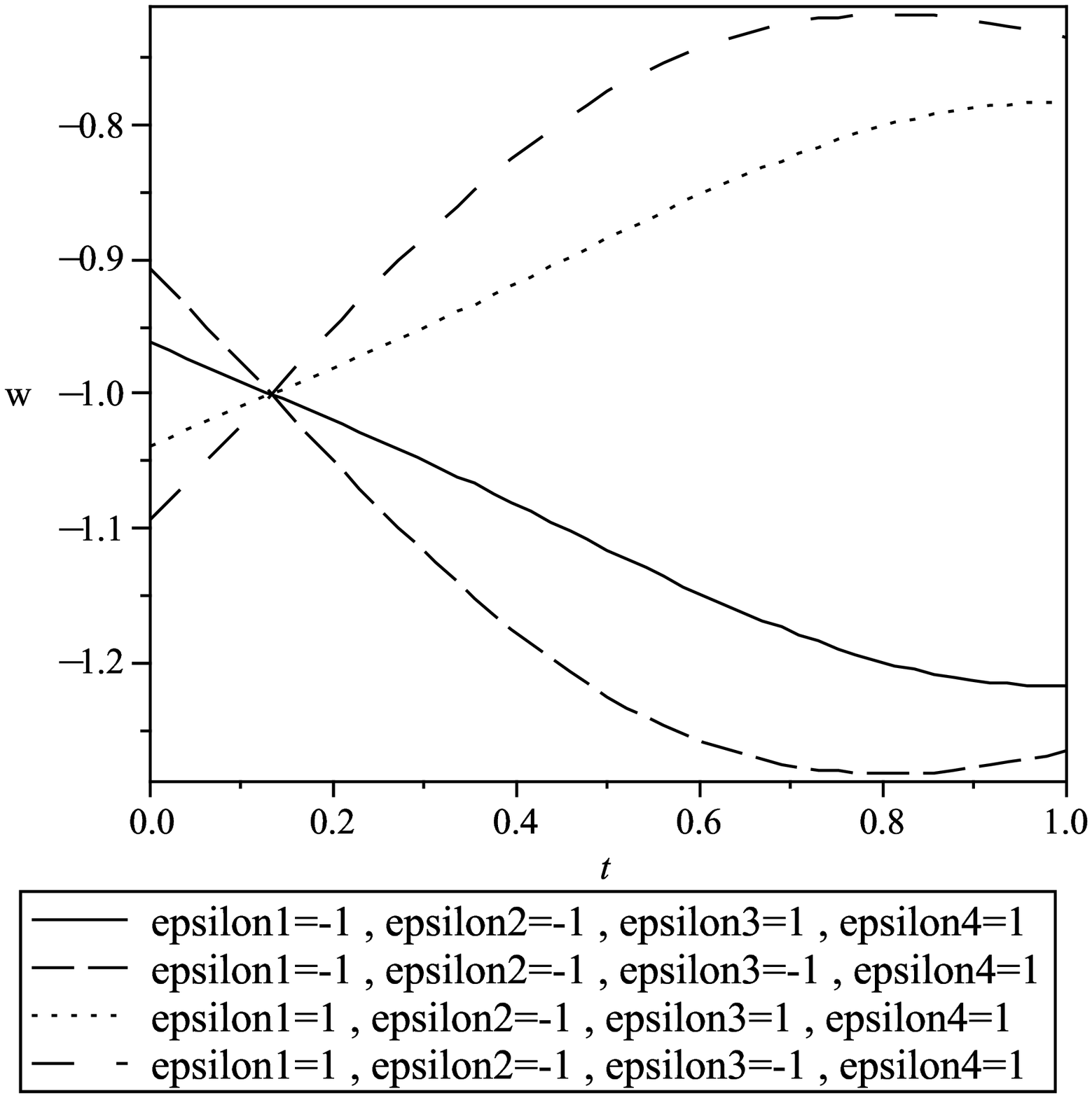} \vspace{-11 cm}  \caption{\small {Crossing the phantom
divide line.}\hspace{-12cm}}
  \vspace{10cm}

\includegraphics{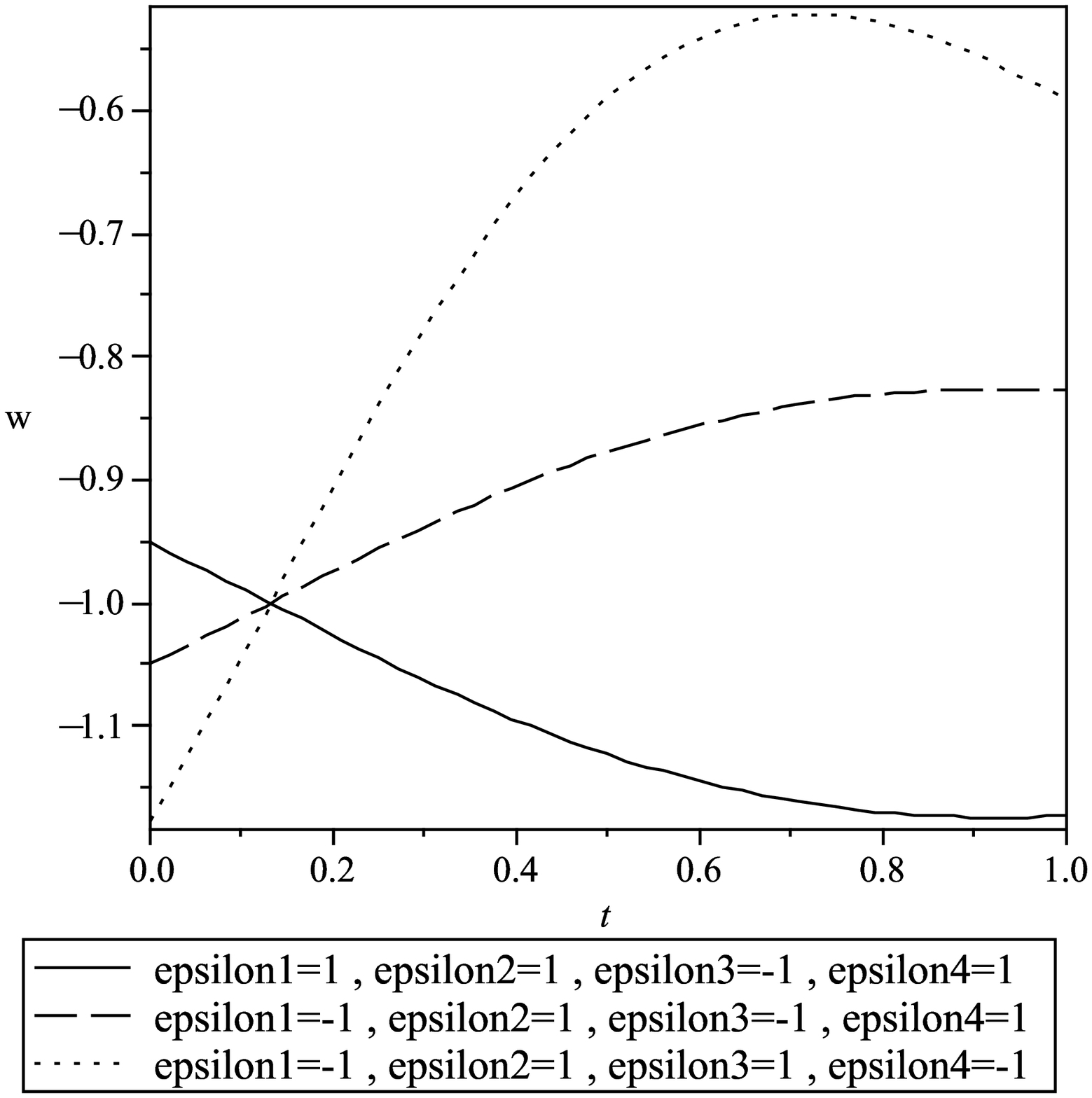}

\vspace{1 cm}
 \caption{\small {Crossing the phantom
divide line. }\hspace{11cm}}

\begin{center}
\includegraphics{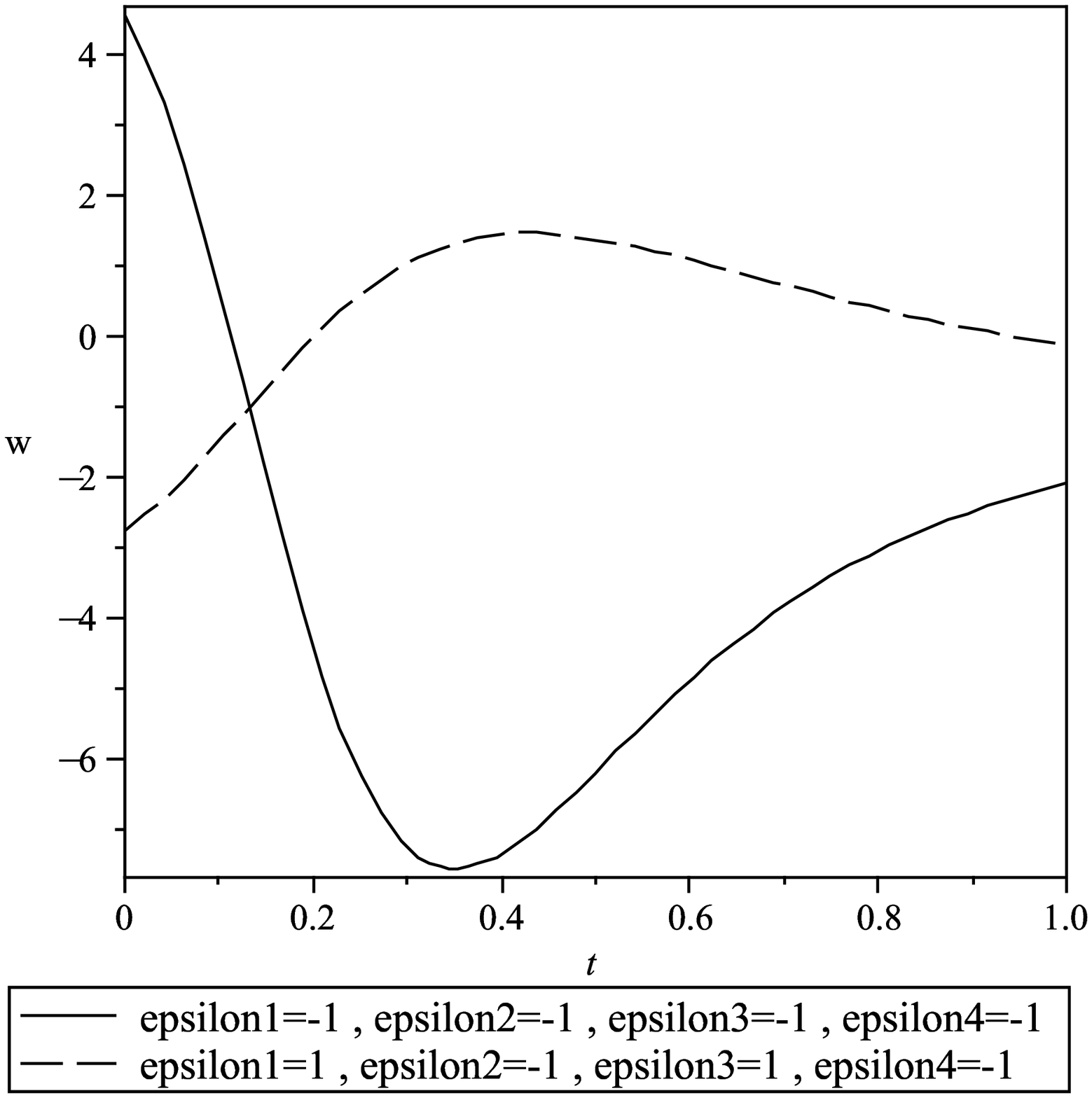}
\end{center}
\vspace{-2.5 cm}
 \caption{\small {Crossing the phantom
divide line.}\hspace{-12cm}}
 \end{figure}
In other hand, one has been discussed in [29-36] about what
conditions lead to a Rip singularity solutions, now we consider, can
this model predict a Rip singularity in context. we note that just
two cases of singularity means Big and Little singularity study in
following.

Therefore, as shown in figures 6-9, we provide suitable conditions
for appear Rip singularity(this situation occur in $\epsilon=+1$
branch). As see in a numerical solution, this situations appear in
the phantom barrier, in this case the equation of state increase
rapidly in finite or infinite time. Hence we can consider the
solution in two major cases as known Little and Big singularities.
In fact we can take a dynamical explication to a case of solutions
that leading to the requirement of Big Rip or Little Rip
singularity. Instead, from this case of solutions we should choose
those requirement which the constraint by observational data. In
other view point, some other types of singularity maybe appears in
our work.
one can easily check other type singularity with suitable conditions in this model. \\
\begin{figure}[htp]
\includegraphics{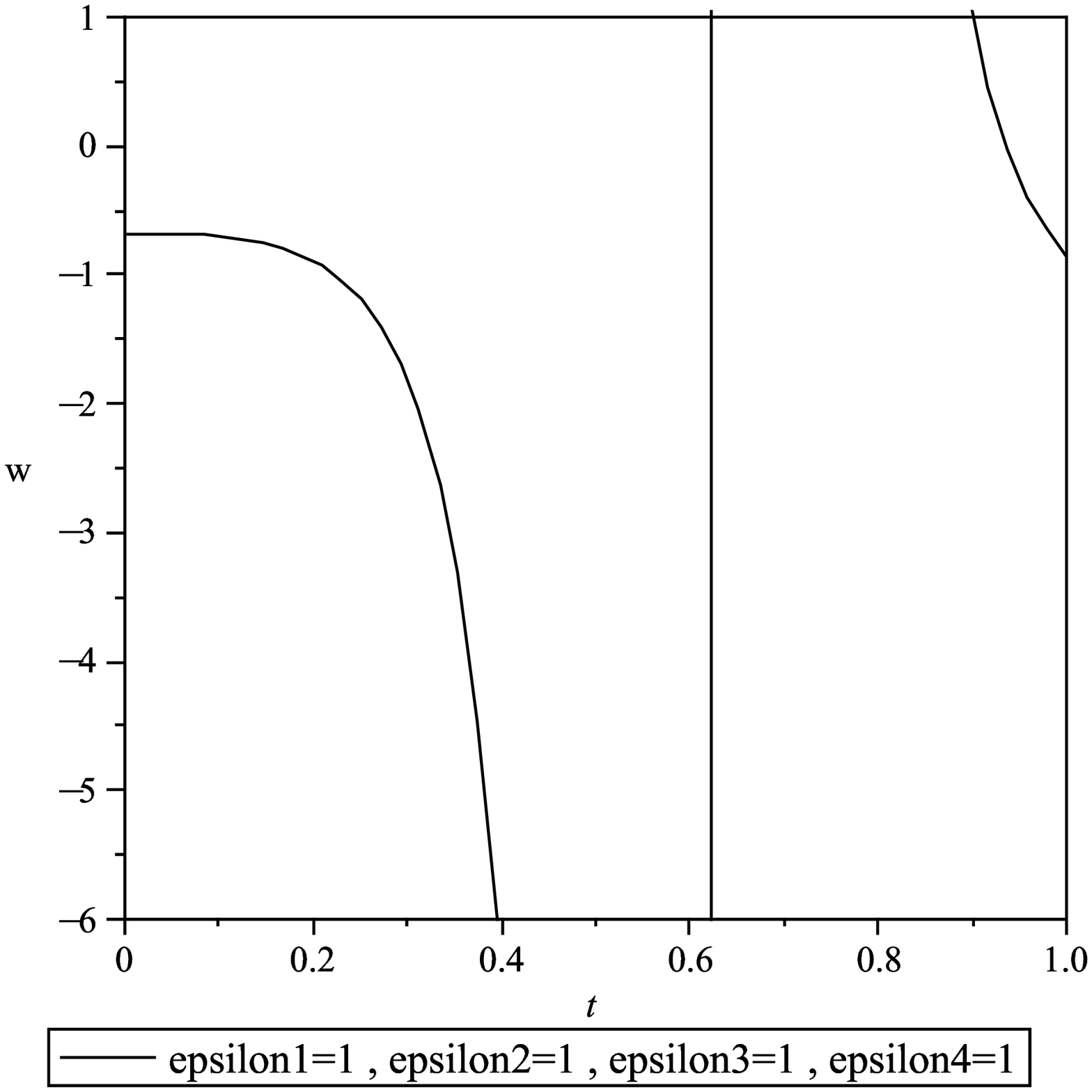} \vspace{7 cm}
 \caption{\small {Rip singularity }
 \hspace{11cm}}
 \vspace{10cm}

\includegraphics{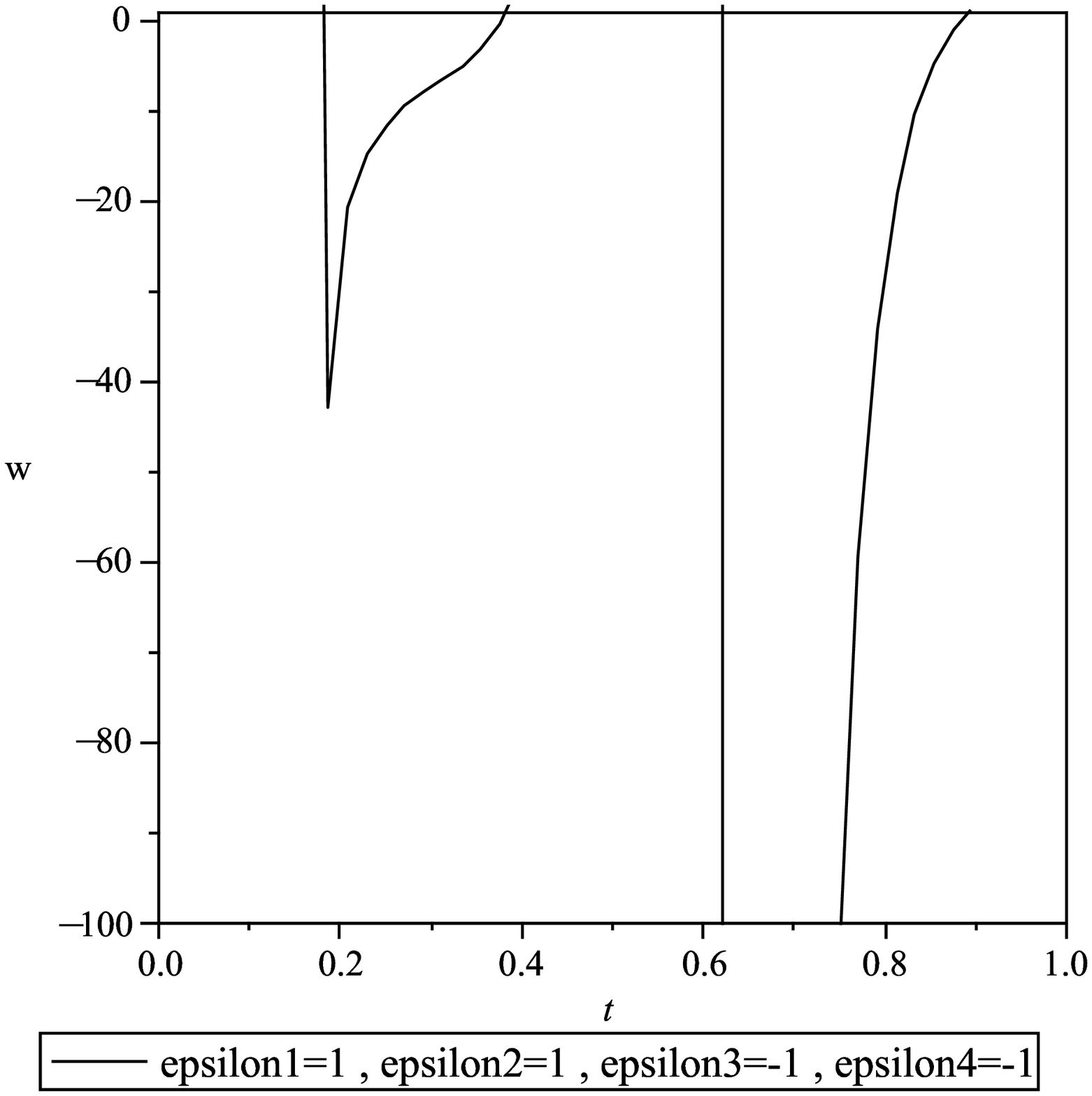} \vspace{-11 cm}  \caption{\small {Rip
singularity}\hspace{-12cm}}
  \vspace{10cm}

\includegraphics{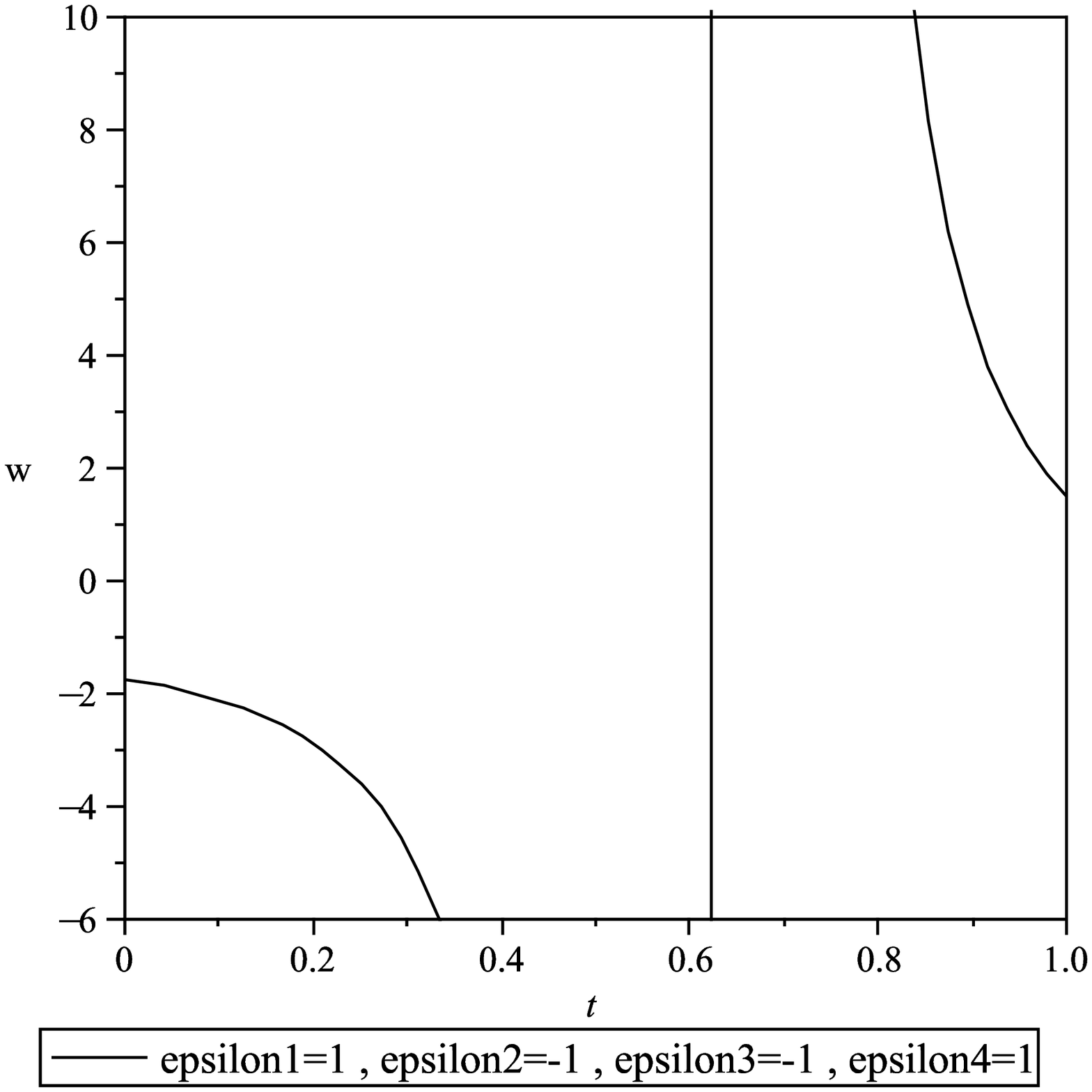}

\vspace{1 cm}
 \caption{\small {Rip singularity }\hspace{11cm}}

\begin{center}
\includegraphics{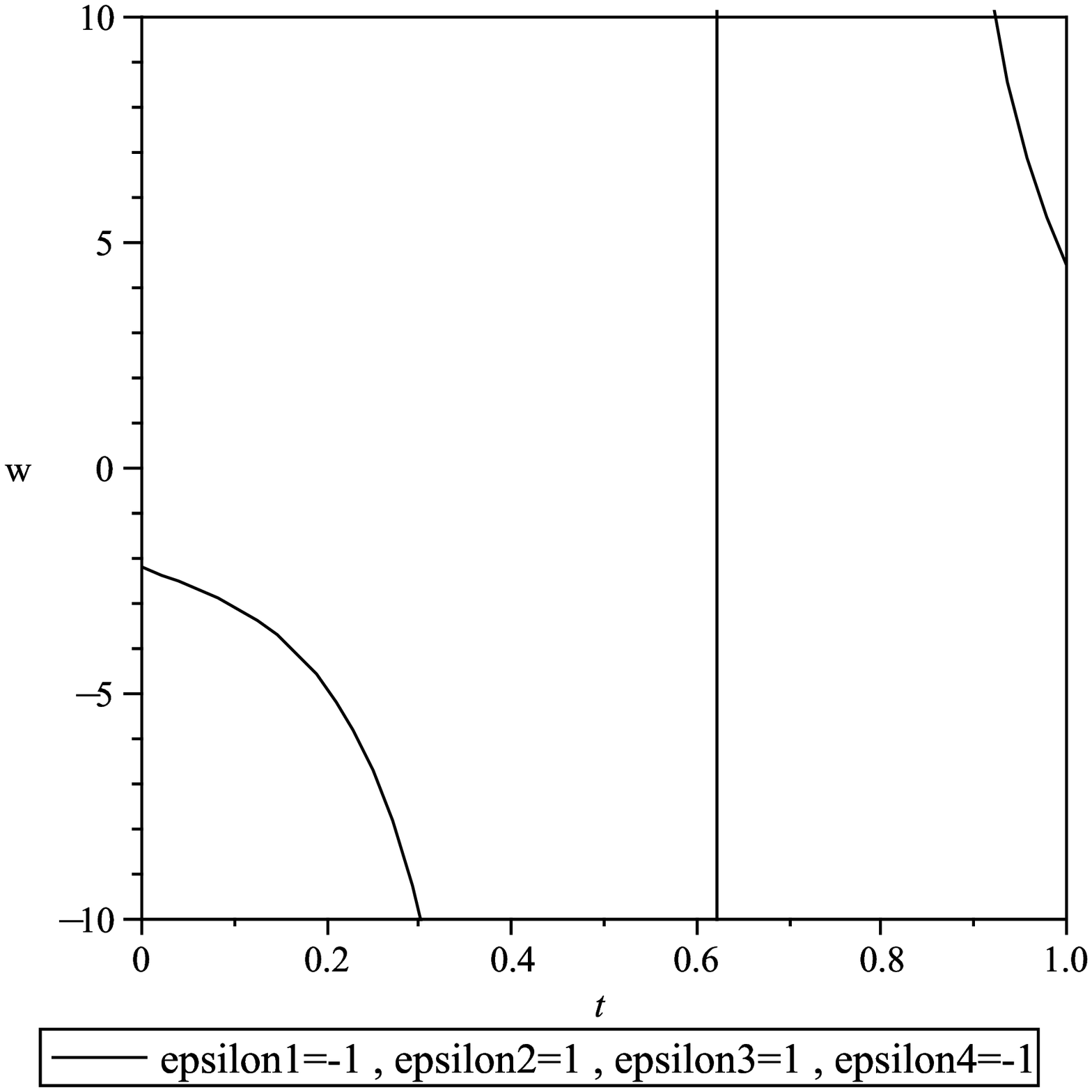}
\end{center}
\vspace{-2.5 cm}
 \caption{\small {Rip singularity}\hspace{-12cm}}
 \end{figure}
\begin{table}
\begin{center}
\caption{Considering Crossing phantom divided line and Rip
singularity in equation of state in equation (36) on all branch for
$\epsilon=-1$ (middle) and $\epsilon=+1$ (right) in equation (21)
for potential of the form \, $V(\phi)=\lambda'\phi^2$} \vspace{0.5
cm}
\begin{tabular}{|c|c|c|c|c|c|c|c|}
  \hline
  \hline $\epsilon_1~ \epsilon_2~ \epsilon_3~ \epsilon_4$& Crossing PDL & Rip singularity \\
  \hline + + + + & Yes & Big\\
  \hline + + + - & Yes & Big\\
  \hline + + - - & Yes & Little\\
  \hline + - - - & Yes & Big\\
  \hline - - - - & Yes & normal\\
  \hline - + + + & Yes & Big\\
  \hline - - + + & Yes & Big\\
  \hline - - - + & Yes & normal\\
  \hline + - + + & Yes & crossing PDL\\
  \hline + - - + & Yes & Big\\
  \hline + + - + & Yes & normal\\
  \hline + - + - & Yes & Big\\
  \hline - + - + & Yes & Big\\
  \hline - + + - & Yes & Big\\
     \hline
\end{tabular}
\end{center}
\end{table}
Now we should describe what case of equation of state is studied for
determine the character of mixed fluids. Typically in this model, we
use three energy-momentum contents, 1: The ordinary matter 2: Dark
energy describe by the scalar field 3: energy-momentum content which
determine by a Lorentz violation vector field. Here we consider that
ordinary matter has partial part of entire energy-momentum content.
But for other contents, in our model it is possible to take the
"trigger mechanism" to describe dynamical equation of state. i.e. we
suppose that scalar- vector-tensor theory with LIV acts similar a
hybrid inflation models. In this regard, scaler and vector field
have the roles of inflaton and the "waterfall" field. Hence, we can
fine-tuning parameters of model for best fit with observational data
[10,25]. It is acceptable to expect that one of them suddenly
dominate and we have a cosmological stage, for example, inflation
phase or acceleration phase and etc. Note that an important result
in this context is crossing of phantom divide barrier PDL that
determine with a single minimal coupling scalar field and by a
Lorentz violating vector field, if considering suitable fine tuning
of model parameters.
\section{Summary}
We have consider a possible violation of Lorentz invariance in a DGP
Brane cosmology. Also we have shown that by a suitable choice of
parameter space, it is possible to have crossing phantom divided
barrier (PDL) and it can predicts Rip singularity in LIV context. We
used an interactive picture, a minimally coupled scalar field and a
Lorentz violating vector field can lead to the phantom phase. The
comparison of our results with cosmology of 4D non-minimal vector
theories show same aspects but with non minimal coupling in vector
term, for example, explain the late time acceleration of the
Universe[26,27,28]. Instead, those models have been shown that a non
minimally vector field can be a candidate for dark energy[28] just
like our model.

In other view point, in this framework, there is the possibility of
a Rip singularity by suitable tuning in the parameters. As shown in
Figs. 6-9, we have studied some type of solutions called Little or
Big Rip, that maybe occur in structures of the Universe, according
to the increasing of the dark energy density. However, In correct
choice of parameters maybe other types of singularity appear in this
model. Finally we emphasize that we have used a model of a scalar
field with Lorentz violation to describe some cosmological aspects.
We believe this model can perform some fascinating predictions for observations.\\

{\bf Acknowledgement}

We would like to thank an anonymous referee for his/her highly
valuable comments.

\end{document}